\documentclass[12pt]{book}

\usepackage[dvips]{graphicx,color}
\usepackage{makeidx,phisics,cosmology}

\makeauthorindex
\makeindex

\BookTitle{Frontier in Astroparticle Physics and Cosmology}
\CopyRight{\copyright 2004 by Universal Academy Press, Inc.}

\begin{document}

\BookTitle{\itshape Frontier in Astroparticle Physics and Cosmology}
\CopyRight{\copyright 2004 by Universal Academy Press, Inc.}
\pagenumbering{arabic}

\chapter{The Origin of E+A Galaxies
}

\author{%
Tomotsugu GOTO and the SDSS Collaboration\\
{\it Department of Astronomy, School of Science, The University of Tokyo, 7-3-1 Hongo, Bunkyo-ku, Tokyo 113-0033, Japan}
}
%
%
\AuthorContents{T.\ Goto} 

\AuthorIndex{Goto}{T.} 

%
%

\section{E+A Mystery}

 E+A galaxies are galaxies with strong Balmer absorption lines and with no
 significant emission in [OII] or H$\alpha$ lines.  
  The existence of strong Balmer absorption lines shows 
 that these galaxies have experienced a starburst within the last 1 Gyr.  At
 the same time, these galaxies do not show any sign of on-going star 
 formation as indicated by the lack of [OII] or H$\alpha$ emission lines.  
   Therefore, E+A galaxies are interpreted as a post-starburst galaxy ---
 a galaxy which truncated starburst suddenly. 
  However, for more than 20 years, the origin of this post-starburst
 phenomena remained a mystery. 
 Using the 100,000 galaxy spectra of the Sloan Digital Sky Survey,
 we tested three popular explanations to the E+A phenomenon.

\section{Are E+As Created by a Cluster-Related Mechanism?}

  At first, E+A galaxies were found in cluster regions. Therefore,
  cluster related physical mechanisms have been thought to be
  responsible for the E+A phenomena. However, in the left panel of
  Fig. 1, E+A galaxies show the same local galaxy density distribution
  as field (all) galaxies. These numerous field E+A galaxies
  cannot be explained with a cluster related physical mechanism.  

\section{Are E+As Dusty Starbursts?}

     Another possible explanation for E+A phenomena is a dusty starburst, where
    E+A galaxies are actually star-forming, but emission 
    lines are invisible in optical wavelengths due to the heavy
   obscuration by dust. 
    However, in the middle panel of Fig.1, all E+As have $r-K<1.1$, where
    the theoretical models predict  $r-K \sim 1.5$ for dusty starburst galaxies. In
    the left panel of Fig.1, no E+As show any sign of starburst in radio
    estimated (dust-free) star formation rate derived from the FIRST data.

\section{Are E+As Meger/Interaction Origin?}

 Merger/Interaction between two galaxies also can induce star
 formation. In Table 1, we counted number of accompanying galaxies
 within 50 and 75 kpc around E+A galaxies and compared those with
 randomly selected normal galaxies. We use galaxies with $-23<Mr<-19.5$
 and statistically subtract background counts using the global galaxy
 number counts in the SDSS data. As shown in the table, E+As have more
 accompanying galaxies than normal galaxies. Especially E+As with the
 strongest H$\delta$ absorption (H$\delta$ EW $>7\AA$) have 8 times more
 accompanying galaxies within 50 kpc. Since this excess in the number of
 accompanying galaxies is significant, we conclude that E+As are most
 likely to have a merger/interaction origin \cite{PhD}\cite{PASJ}.

\begin{table}[h]
\caption{
 Number of accompanying galaxies around all the E+A galaxies and E+As
 with H$\delta$ EW$>$7 are compared with that of 1000
 randomly picked galaxies. We use galaxies 
 with -23.0$<Mr^*<$-19.5 after $k$-correction within the radius of 50 and 75
 kpc. Fore/background galaxy number counts are  statistically subtracted. 
}\label{tab:ea2_nearby_galaxies}
\begin{tabular}{lll}
\hline
 Sample  & $N_{nearby\ galaxy}$ within 50 kpc  & $N_{nearby\ galaxy}$ within 75 kpc \\
\hline
\hline
 Random                         &   0.03$\pm$0.01 & 0.16$\pm$0.01  \\ 
 All E+As                       &   0.12$\pm$0.03 & 0.26$\pm$0.04  \\ 
 E+A with H$\delta$ EW$>$7 \AA  &   0.24$\pm$0.09 & 0.40$\pm$0.12  \\ 
\hline
\end{tabular}
\end{table}

\begin{figure}[t]
  \begin{center}
    \includegraphics[height=10pc]{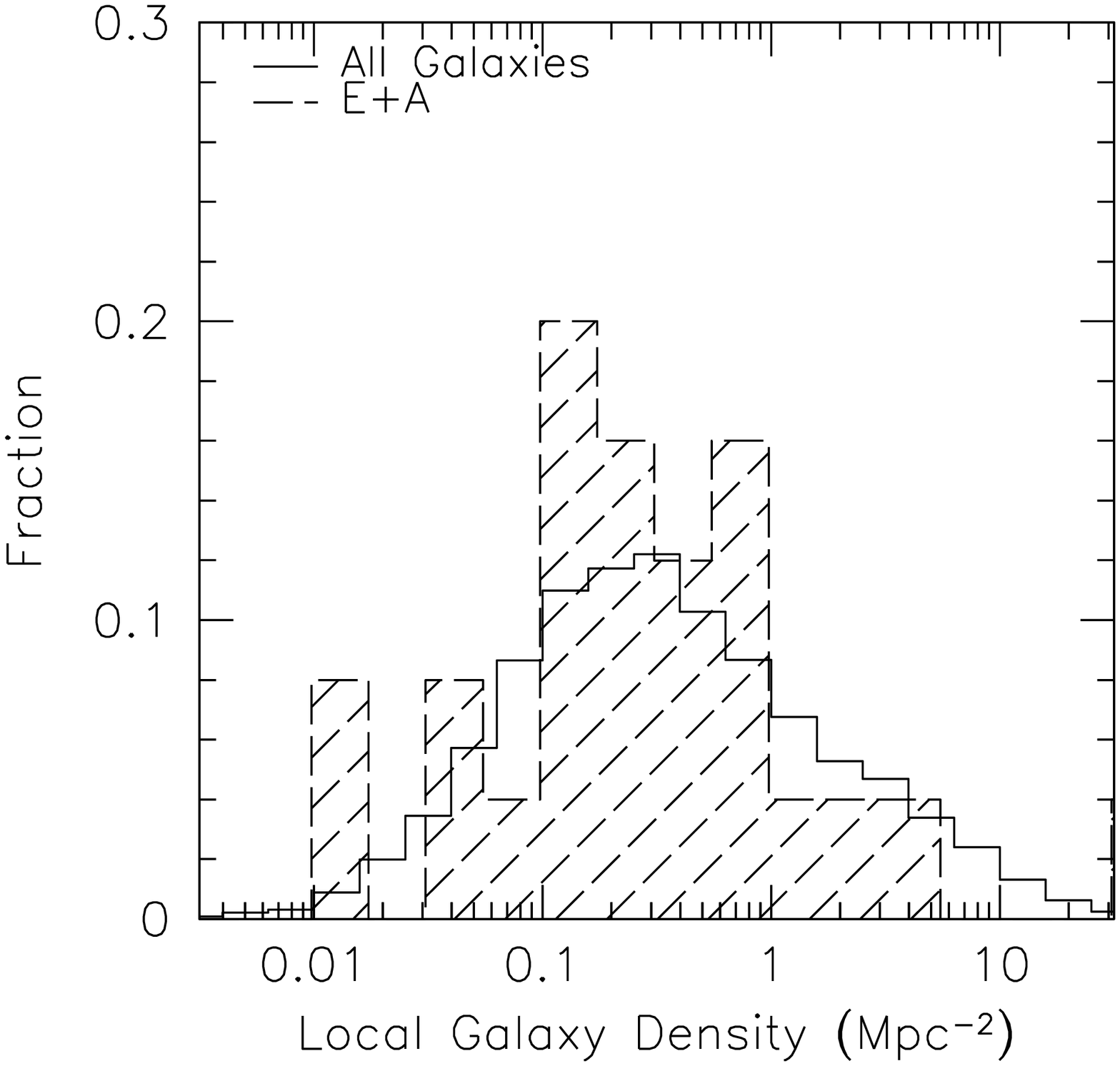}
    \includegraphics[height=10pc]{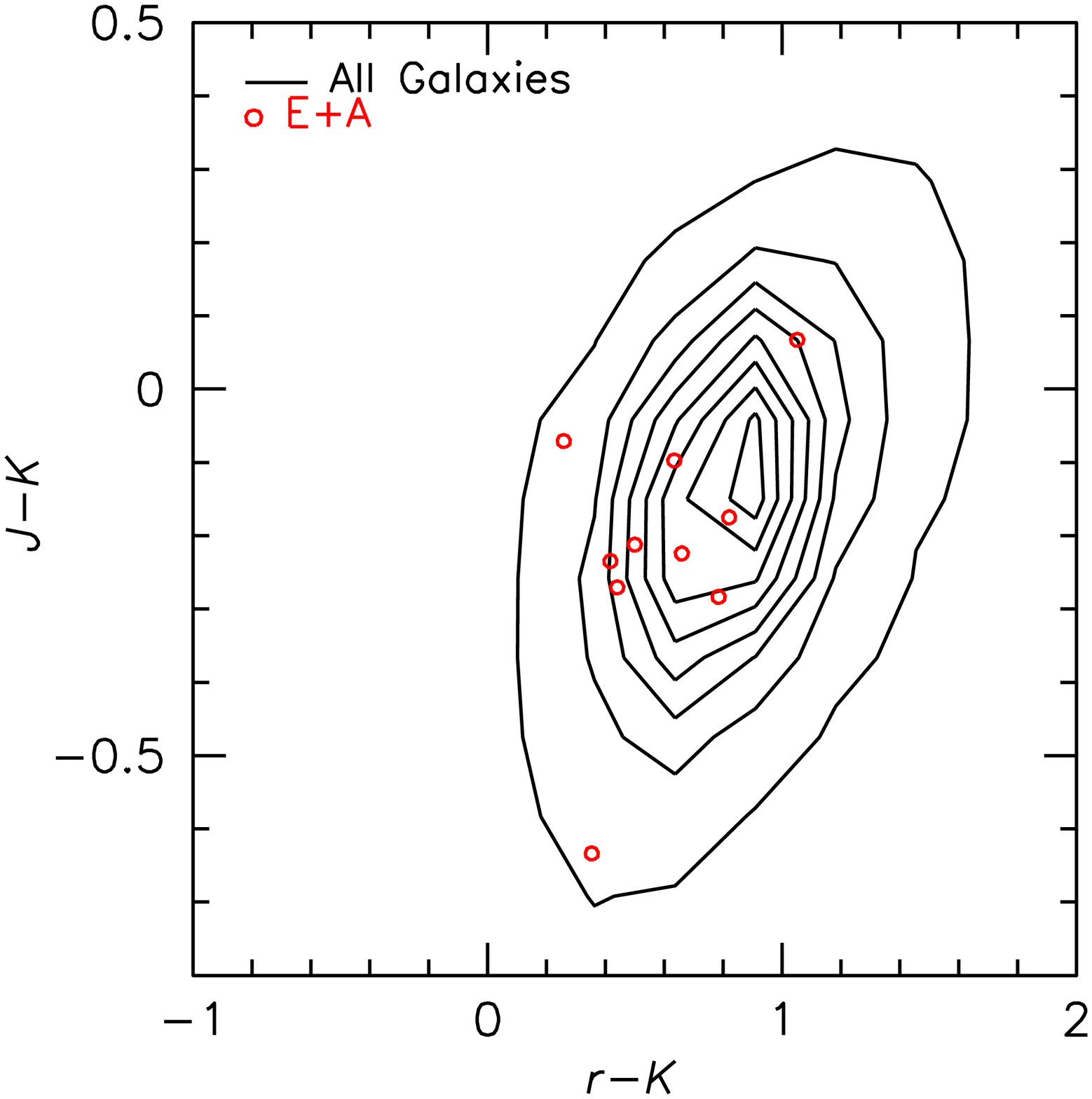}
    \includegraphics[height=10pc]{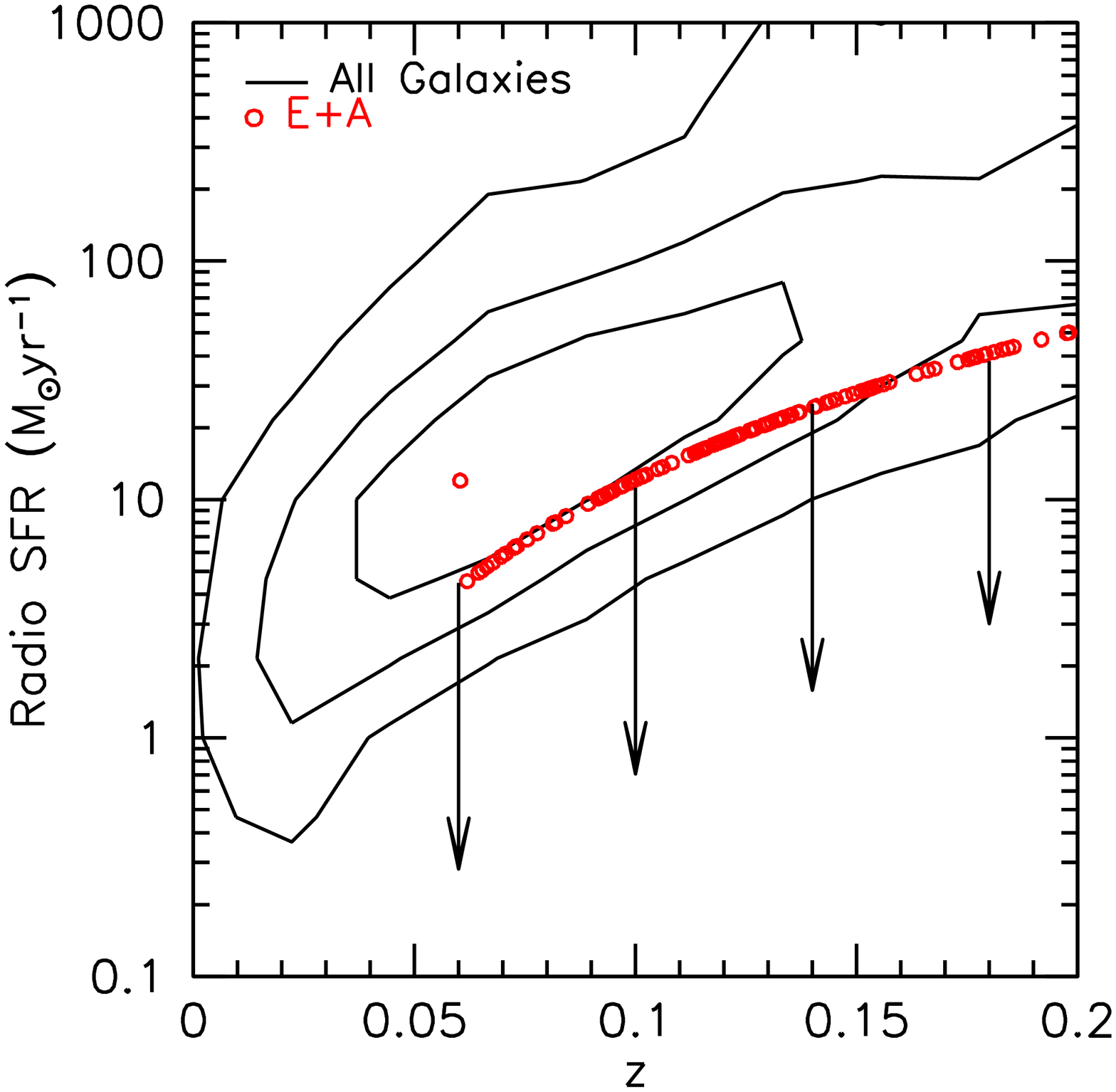}
  \end{center}
  \caption{(left) Local galaxy density distribution for E+As and all
 galaxies. (middle) SDSS-2MASS optical-near infrared color distribution.
 (right) Radio estimated star formation rate derived from the FIRST data.}
\end{figure}

\end{document}